\documentclass[debug]{rmaa}

\usepackage{paralist}

\usepackage{psfrag,color}

\usepackage[latin1]{inputenc}
\usepackage{epstopdf}

\title{ Infrared characteristics of the BIS catalog objects  }

\author{
S. Gaudenzi,\altaffilmark{1} 
R. Nesci,\altaffilmark{1} 
C. Rossi,\altaffilmark{2,3}
 S. Sclavi,\altaffilmark{3} 
 K. S. Gigoyan,\altaffilmark{4} 
and  A. M. Mickaelian,\altaffilmark{4} 
  }

\altaffiltext{1}{INAF/IAPS, Roma Italy.}
\altaffiltext{2}{INAF/ Osservatorio di Monte Porzio, Roma, Italy.}
\altaffiltext{3}{Universit\`aLa Sapienza Roma, Italy.}
\altaffiltext{4}{ Ambartsumian Byurakan Astrophysical Observatory (BAO). }

\shortauthor{ Gaudenzi, Nesci \& al.}
\shorttitle{ BIS sources : infrared results.}

\fulladdresses{
\item S.Gaudenzi and R.Nesci: INAF/IAPS, via Fosso del Cavaliere 100, 00133 Roma, Italy  (silvia.gaudenzi@iaps.inaf.it,  roberto.nesci@iaps.inaf.it ).
\item C. Rossi: INAF/Osservatorio Astronomico di Roma, Via Frascati 33, 00040, Monte Porzio Catone (RM), Italy (corinne.rossi@uniroma1.it).
\item S.Sclavi and C. Rossi: Dipartimento di Fisica, Universit\`a La Sapienza,
Piazzale Aldo Moro 3, 00185 Roma, Italy.
\item K.  S.Gigoyan and A.M.Mickaelian V. A. Ambartsumian Byurakan Astrophysical Observatory (BAO) and Isaac Newton Institute of Chile, Armenian Branch, Byurakan 0213, Aragatzotn province, Armenia.
  }

\listofauthors{ S.Gaudenzi, R.Nesci  \& al.}
\indexauthor{Gaudenzi, S.}
\indexauthor{Nesci, R.}
\indexauthor{Rossi, C.}
\indexauthor{Sclavi, S.}
\indexauthor{K. S. Gigoyan.}
\indexauthor{Mickaelian, A.M.}

\abstract{
We studied several color-color infrared diagrams of the 276 late type stars of the BIS catalog. For 95 of these stars we derived spectral classification from our slit spectroscopy. From the 2MASS color diagram we derive that none of the sample stars is a dwarf. The $WISE$ 3.4-12 vs 12-22 plot is the best to discriminate the variability type (Mira, irregular and not-variable), as well as Carbon stars: IRAS colors are less useful due to the poor quality of the data for most of the BIS stars. Mixed plots involving 2MASS, $AKARI$ and $WISE$ were also explored: the 2MASS-AKARI J-[S09] vs [S09]-[L18] plot is efficient to discriminate Carbon and Mira stars. From the color plots and  our spectroscopy we can statistically predict that in whole BIS catalog no other dusty Carbon star is present, while a few Miras and CH or R Carbon stars could be discovered. Only about 15\% of the BIS stars are of early M type.
}

\resumen{Estudiamos diagramas color-color en el infrarrojo de 276 estrellas del tipo tardío del catálogo BIS. Hemos usado archivos de 2MASS, $WISE$, AKARI e IRAS. Para 95 de estas estrellas, se encuentran disponibles las clasificaciones espectrales de nuestra espectroscopía de rendija y las curvas de luz en el óptico de los archivos de NVSV y Catalina.
Con el diagrama de color de 2MASS derivamos que ninguna de las estrellas de la muestra es una estrella enana. La grafica 3.4-12 vs 12-22 de $WISE$ es la mejor opción para diferenciar el tipo de variabilidad (Mira, irregular y no variable), así como las estrellas de Carbón: También es útil el diagrama 3.4-4.6 vs 4.6-22. Los colores de IRAS son menos útiles debido a la pobre calidad de los datos de la mayoría de estrellas BIS. Se exploraron diversas graficas de datos de 2MASS, AKARI y $WISE$ y se encontraron correlaciones aceptables entre magnitudes de pasabandas similares de diferentes instrumentos. La gráfica 2MASS-AKARI J-[S09] vs [S09]-[L18] es también eficiente para diferenciar las estrellas de Carbón de las estrellas Mira. 
A partir de las graficas de colores y de nuestra espectroscopía disponible podemos predecir estadísticamente que en todo el catálogo BIS no existe otra estrella de Carbón con polvo, mientras que algunas estrellas Miras y estrellas de Carbón CH o R pueden ser descubiertas. Solo alrededor del 15\% de las estrellas BIS son del tipo temprano M.
}

\addkeyword{}
\addkeyword{Stars: late type}
\addkeyword{Stars: variables}
\addkeyword{Stars: Infrared sources}

\begin{document}
\maketitle

\section{Introduction: Byurakan -- IRAS Stars catalog}  \label{Intro}

The first major mission covering mid- and far- InfraRed (IR) was the InfraRed Astronomical Satellite (IRAS) in 1983, which surveyed about  96\% of the  sky in four bands centred  at 12, 25, 60, and 100$\mu$m. 
Two general catalogs were produced, IRAS Point Source Catalog  \citet{PSC88}   and IRAS Faint Source Catalog (FSC; Moshir et al. 1989, more useful for extragalactic studies). 

A recent work by     \citet{Abr15} made a cross-check of the PSC and FSC catalogs using a new cross-correlation technique and establishing correspondence between these two catalogs
 and cross-correlation with more recent IR missions: 
  2MASS (1.25, 1.65  and 2.17 $\mu$m, \citealp{Cut03, skrut06}),
  $WISE$   (3.4, 4.6, 12 and 22~$\mu$m,   \citealp{Cut13}), 
  and $AKARI$  (9 and 18 $\mu$m,  IRC, \citealp{Ishi10}; FIS, \citealp{yama10}).
 In this way accurate positions for IRAS sources and fluxes at 17 bands from $J,H,K$  to far-IR
 were provided.
 
InfraRed data available from several archives, make it possible to build many colour-magnitude and colour-colour diagrams for the study of the distribution of objects
  and for the first characterisation of these sources, finding possible grouping, as well as unique objects.
   A number of such statistical works have been published by  several groups  (e.g. \citealp{ita09, ita10, kato12,  murata14, nikutta14, Tu13, tisse13}).
Accurate optical identifications of IR sources are important for further studies and the comparison of the optical and IR characteristics    of the objects.
 In the PSC Vizier catalog (II/125), 142,228 associations are given (often several associations for the same IRAS source) accomplished by cross-correlations with several optical catalogs. Similarly, in the  FSC Vizier catalog (II/156A), 235,935 are given. However, these associations should be verified and they may be used for preliminary studies only.
  Therefore, a number of works were devoted to optical identifications of IRAS sources using positions and fluxes, very often pre-selecting by types of objects and/or galactic latitudes.
    Among these projects, one was conducted by \citet{mic1995} 
    in the Byurakan Astrophysical Observatory  using DSS images and low-dispersion spectra from the Markarian Survey  in the area with $\delta$ $>$ +61$^o$ and galactic latitude b $>$ +15$^o$.
     Identifications were accomplished for all 1578 IRAS PSC sources present in the area,  without any discrimination  in order to keep the completeness. Later on stars and galaxies were distinguished as Byurakan IR Stars (BIS) and Byurakan IR Galaxies (BIG). Several lists of IRAS stellar sources  were published by Mickaelian and Gigoyan between 1997 and 2001; the resulting catalogs provided optical and IR data for 276 IRAS BIS objects (see \citealp{mic06}).

Aiming at clarifying the nature of the stars included in the BIS catalog we have started a systematic collection of all the information available in literature and acquisition of new data. 
 In a companion paper  (\citealp{gaud17}, hereafter G17) we  have studied  optical spectroscopic  and variability  characteristics of a subsample of 94 BIS targets. The present work is dedicated to the IR characteristics of the whole BIS catalog (275 stars, being BIS\,028 a planetary nebula).
We will use all the available data from the following catalogs:
 2MASS, $WISE$, $AKARI$,  and $IRAS$. 
We will also use the information gained from our optically studied subsample  to interpret the infrared color-color plots and make statistical predictions on the overall stellar content of the whole BIS sample.

 \section{Our targets}  \label{OLTS}

The  BIS catalog  contains candidates of several spectral types: M$-$type and carbon$-$type stars, Mira$-$type and Semi$-$Regular (SR) variables, OH and SiO sources, and unknown sources surrounded by thick circumstellar shells.
Most of the sources  are poorly studied both from the photometric and spectroscopic point of view and only a few of them are classified as variables in the GCVS \citep{Sam17} or VSX \citep{watson16}  catalogs.
As described in G17,
we have  revised the spectral types,  the variability behavior  and determined a number of physical parameters for 1/3 of the BIS studied sources.
 Here  we briefly remember the  results
  relevant for the  discussion of the  infrared  color$-$color diagrams described  below.

{\bf Variability class~(1): Mira type  variables} 
{\it BIS\,007;  116; 133 (IY~Dra); 196;  267 }. 
 With the exception of BIS\,116  these are late type Miras, varying of one subtype depending on the phase.

{\bf Variability class~(2): semi-regular, irregular variables} 
  {\it  BIS\,001; 002;  003;  004;  006; 014; 015; 032; 036; 037; 038; 039; 043 (KP~Cam); 088; 103; 104; 106; 120; 122; 123; 126; 132; 136; 138; 142; 145; 154; 156; 167; 168; 170; 173; 184; 198; 200; 207; 209; 211; 212; 213; 214; 216; 219; 226; 264; 271; 276;  }.
We have grouped here  semiregular and irregular variables  with visual amplitude larger than 0.3 magnitudes. The spectral types are  from M5  to later ones.

{\bf Variability class~(2/3): small amplitude variability} 
{\it  BIS\,010;  107;  113;  172;  174;  201;  224;  255;  275;  285}.
 These stars show small ( $<$ 0.3 magnitudes), a-periodical variations;    with the exception of BIS\,113,~(M7)  the spectral types are in the range M3 - M5.
 
{\bf Variability class~(3): very stable stars}
{\it BIS\,034;  044;  067;  087;  099;  102;  110;  137; 143;  155; 194; 197; 199;  203; 210;  228;  247; 248;  256;  260}. The level of  optical variation is never larger than 0.2 magnitudes;
 the spectral types are in the range M0 - M4.

{\bf Carbon stars: }
{ \it BIS\,036; 184 (HP~Cam); 222; 194}: all are N giants.
The first three stars are Irregular variables, embedded in a  dusty envelope;  BIS\,194 is stable. 


 \section{Color$-$color diagrams }   \label{DA}
 
    There is a wide literature describing the use of homogeneous and non homogeneous data sets combined in several ways with the aim to distinguish  different types  of sources in color$-$color diagrams.
During our work on the infrared data we have applied already tested combinations and have investigated the possibility to find new efficient discriminators for late type stars.
In what follows we will describe some of the diagrams that we consider  the most successful and  will mention  others also useful  to our purpose.
We will  discuss and compare our results with those already known from the most popular diagrams. 

 When   non homogeneous data are  considered,  variable stars are most probably observed in 
  different photometric  phases  but we have verified that
  this does not affect their position in  color$-$color diagrams ( see \S~\ref{akari}). 
  This can be explained,  because
 the  variability amplitude decreases with the increasing wavelength, being  seldom larger than  0.5 mag  already at $K$ waveband. 

As stated above,  the diagrams  include  the sources of the BIS catalog not studied in G17.  They are plotted as open triangles in all figures; from the NSVS light curve    \citep{woz04a}, 54  are  strongly variables  (class 1 or 2 of our classification), 66 are  very stable,
27 show minor oscillations (class 2/3), the remaining are not present in the NSVS database.

  \medskip
\subsection{2MASS  ~~(1.25, 1.65, 2.17$\mu$m)}
\begin{figure}
\centering
\includegraphics[width=9cm,angle=0]{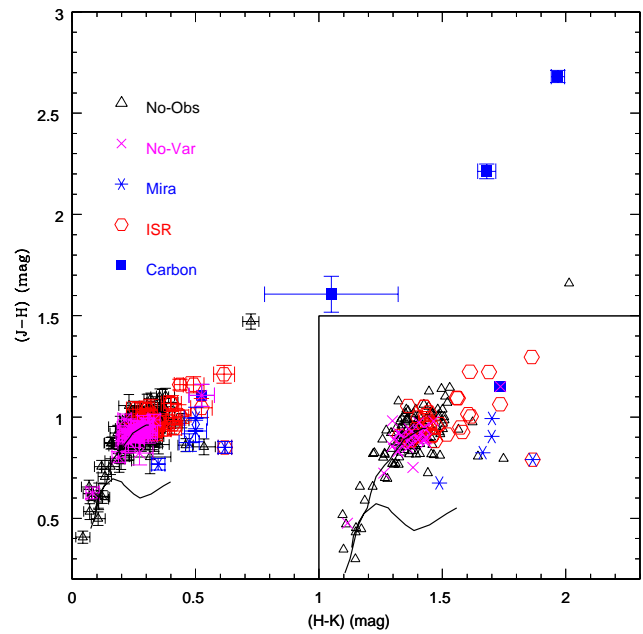}
\caption{$J-H$ {\it vs} $H-K$ color$-$color diagram of  all the BIS sources with 2MASS magnitudes.  
The inset is a zoom of the region of the lower left corner. The symbols indicate our variability classes: magenta crosses= not variable; blue asterisk= M Mira; red polygon=Irregular-Semi-Regular; filled blue squares=  carbon stars; black open triangles= stars not yet spectroscopically observed.  Error bars are not indicated in the insert. The two lines show the locus of the giants(upper)  and dwarfs (lower) from \citet{bessel88}. The separation starts at about spectral type K9
}
\label{fig1}
\end{figure}

 In the near infrared  the  data for almost all our targets  are available in the 2MASS catalog.
To discriminate dwarf/giant luminosity class, we used the traditional 
color$-$color plot $J-H$ {\it vs} $H-K$ \citep{bessel88, lege02, reid02, kirk99, cruz03}. 
We have transformed the 2MASS magnitudes into the Bessel \& Brett system using the formulae given in the Explanatory Supplement to the 2MASS Second Incremental Data Release
\footnote{http://www.ipac.caltech.edu/2mass/releases/second/doc/  ~sec6\_3.html} and in  \citet{carp01}.
The results  presented in Fig.~\ref{fig1}  show that among the stars studied in G17 none is located in the dwarfs region, namely $(J-H) <$ 0.7~mag and ($H-K$)$>$ 0.15~mag, while the colors are all typical for giants and AGB stars (see  Figure~A3  and Tables II and  III of \citealp{bessel88}).
The M stars occupy different regions depending on their spectral types: as expected, the stable stars are in the blue corner.
Among  the  variable stars, all the Miras lie below  the bulk  populated by the irregular variables,
in accordance with the well established results found by the $SAAO$ astronomers in their extensive studies ( e.g. \citealp{feast90}).
In addition to this  result, this plot confirms to be  essentially useful to point out heavily obscured, long period variables. Actually, on the basis of their spectra in G17 we classified BIS\,036, 184 and 222 as N$-$type dust enshrouded stars. 
In the $J-H$ {\it vs} $H-K$ diagram these stars are well separated from the others, and located along the path defining heavily obscured, mass losing, Long$-$Period N$-$type AGB stars \citep{loon97}. 
The colors of the fourth carbon star BIS\,194 place this star among the Semi$-$Regular M variables,  but also nicely aligned with the dusty carbon stars. 

%
%

\medskip
\subsection { WISE ~~(3.4, 4.6, 12, 22 $\mu$m)}
The  $WISE$ magnitudes have a widespread application in  searching for reliable criteria to define occupation regions of the various infrared sources in different types of diagrams. 
The huge amount of detected sources allows statistical approaches but hampers satisfactory results to be obtained if the original data are not  matched with catalogs of known objects and with theoretical models.
 \citet{Tu13} have investigated different color-color diagrams to  efficiently separate C-rich from O-rich stars. According to these authors, even their best diagram  ($W2 -W3$ vs $W1-W2$) did not reach the goal.
 \citet{lian2014}, and  \citet{nikutta14}  have overlapped DUSTY models to observations in different color-color diagrams, among which the same as  \citet{Tu13} and $W1-W2$ vs. $W3-W4$  which gives more convincing results.

 Almost all  the stars of the BIS catalog are  present in the $WISE$  database. 
We have tested all the 15 combinations  of the $WISE$ colors,  looking for a criterion able to distinguish  not only the spectral types, but also the variability classes.
We remember that  from G17  knowing  spectral types and  photometric behaviours, we have obtained satisfactory results in this sense too.

In  Figure 2 of \citet{rossi16} we have shown  two of such diagrams: $W1-W2$ vs  $W3-W4$ and $W1-W2$ vs $W1-W4$.
The first diagram is efficient in separating  dusty carbon from O-rich stars, and the Mira stars lie in a narrow region of the  x axis, clearly isolated above the other variables. 
The not-obscured N stars and the CH stars cannot be  isolated in this diagram. The second diagram is more confused.

\begin{figure}[h]
\centering
\includegraphics[width=12.5cm,angle=0]{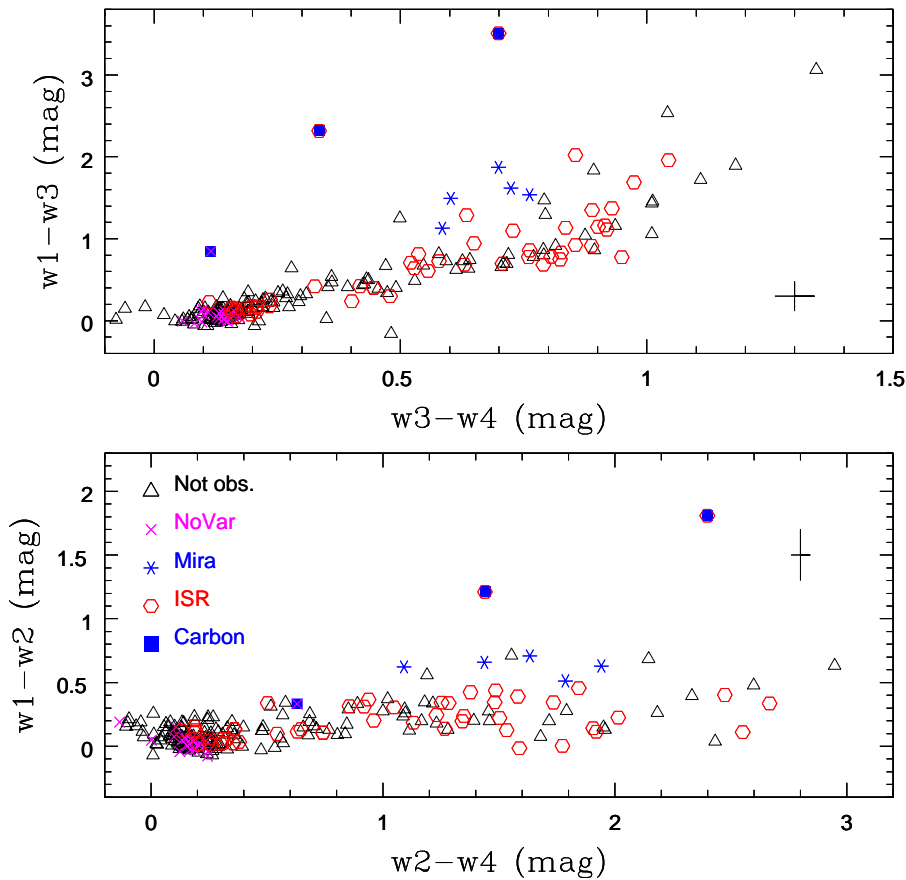}
\caption{ 
 $WISE$ ~color$-$color diagram of the BIS stars: $3.4-12$ ($W1-W3$) {\it vs} $12-22$ ($W3-W4$)  and $3.4-4.6$ ($W1-W2$) {\it vs} $4.6-22$ ($W2-W4$).
 The mean values of the errors are indicated by the two crosses.
( $ \sigma (w1-w2) = 0.19,  \sigma (w2-w4)= 0.091, \sigma (w1-w3)= 0.18,  \sigma (w3-w4)= 0.04 $)
 }
\label {fig2}
\end{figure}

In the present paper  we examine two not yet explored diagrams: the ($W1-W2$) {\it vs} ($W2-W4$) and ($W1-W3$) {\it vs} ($W3-W4$) (see Fig.~\ref{fig2} ).
 In this figure we do not report individual error bars to avoid confusion in the lower part of the diagrams. The mean values are indicated by the two crosses at the corners.
The large $\sigma$~ of $W1-W2$~  and $W1-W3$~ are due to the spread of the uncertainties of $W1$~ having a mean value of 0.18\,mag. 

%
%

In the upper plot the early type, non variable  stars are tightly grouped in the lower left corner
( 0.08$\le W3-W4 \le$ 0.23 and -0.2$\le W2-W4 \le$ 0.3 ), the Semi$-$Regular variables  are distributed along a diagonal strip in the lower part of the  diagram, all the confirmed M~Miras lie above the strip; the  carbon stars, including the naked, stable BIS\,194, are aligned along a separate path in the upper region, far from all the other stars. 
The lower plot is less efficient in separating C-rich from O-rich stars.
BIS\,184  is not reported in these plots because of the large uncertainties of its $WISE$ magnitudes.

\medskip
 \subsection  { IRAS ~~(12, 25, 60 $\mu$m)}
\begin{figure}  [t]
\centering
\includegraphics[width=9cm,angle=0]{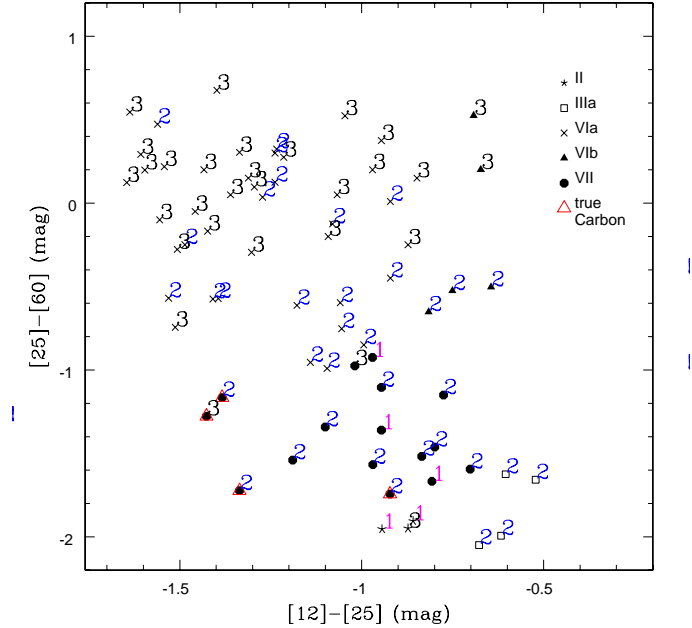}
\caption{IRAS $[12]-[25]$ {\it vs} $[25]-[60]$ color$-$color diagram of our 
target stars. The symbols correspond to the spectral classification  by \citet{veen88}. Spectroscopically confirmed carbon stars are  indicated by red triangles. 
The roman numbers indicate our variability class.
}
\label{fig3}
\end{figure}

The  IRAS database was used by \citet{mic06}, to select the original BIS sample. 
We remember that the data are released with fluxes in Jansky while in the diagrams the magnitudes are normally used. 
We have located our stars in the classical [25]$-$[60] {\it vs} [12]$-$[25] color$-$color plot, 
 where the magnitudes are computed as -2.5Log(Flux) in order to have a direct comparison with Figure 5b of  \citet{veen88}.
In Fig.~\ref{fig3} the different symbols indicate the regions of the classification scheme by \citet{veen88}: 
region II, variables with young O$-$rich shells; 
region IIIa, variables with O$-$rich shells; 
region VIa, non variable stars with relatively cold dust at large distance; 
region VIb, variables with hot dust nearby and cool dust more distant; 
region VII ,  stars with evolved C$-$rich circumstellar shells. 
 To avoind confusion  we  have included only stars  of the BIS catalog having spectral type and variability class known from G17. The pattern does not change adding the remaining stars.
 The numbers indicate the variability class defined above
(1 = Mira type, 2 = semi-regular/irregular variables, 3 =  non variable stars). 
 To evaluate the significance of the result  we must remember that  only quality factors, not absolute uncertainties are reported in the  $IRAS$ database.
  For our stars we remark that  at  60$\mu$m only four stars have  quality factor q=3,  seven q=2  and 59  q=1,  implying that most of the position might  be affected by the low precision of the fluxes.  We have discarded few stars having quality factor 1 already at 25$\mu$m  and being too far from the bulk of the diagram.  

The location of the stars presented in  Fig.~\ref{fig3} is statistically in agreement  with the classification by \citet{veen88} although  this scheme appears not useful to give indication about the  stellar variability of faint sources.
   Only  few stables are  not placed in region VIa, where on the contrary, many strongly variables are also located. Concerning the spectral types,
  the four carbons are all placed in the expected region VII,  occupied by M type variables too and by the stable BIS\,113, having all the  quality factors equal to 1. 
Computing the magnitudes according to the Explanatory Supplement to the Point Source Catalogue   \citep{PSC88}  would only shift  the axes but would not change the relative positions of the sources.   

\medskip
 \subsection { AKARI ~~(9 and 18 $\mu$m) and  composite diagrams} \label{akari}
\begin{figure*}[t]
\centering
\includegraphics[width=12cm, height=12cm, angle=0]{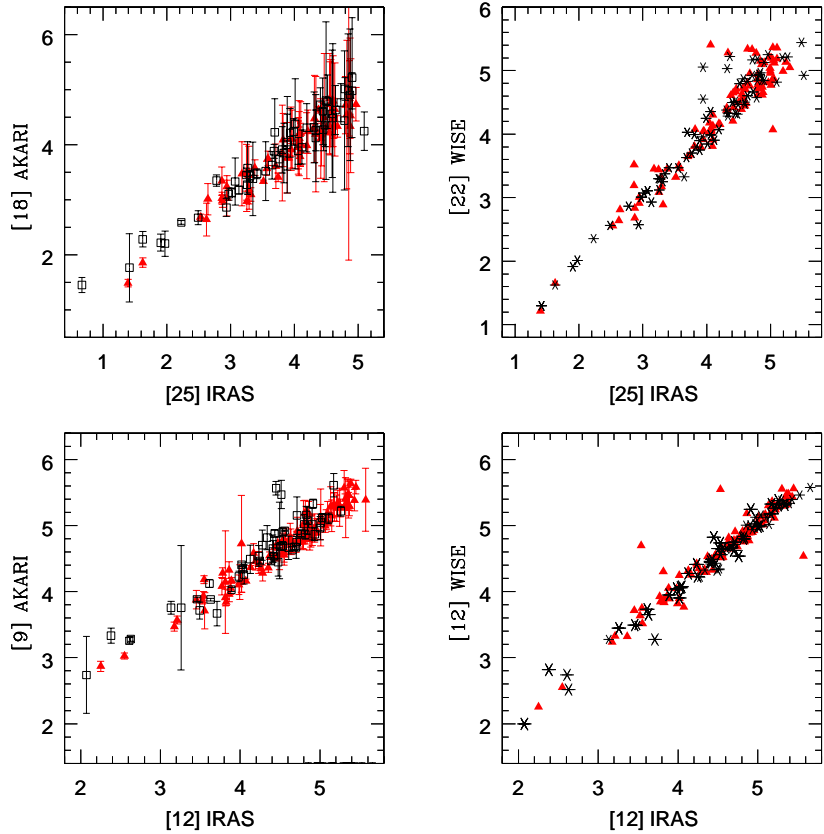}
\caption{  Correlation between  magnitudes  $AKARI - IRAS$ and $WISE - IRAS$. The symbols indicate:  black stars and open squares = stars  present in G17; red filled triangles = stars not present in G17.
 In the right panel the  mean errors are 0.016 for $W3$ and 0.02 for $W4$;  $IRAS$ only provides quality factor.          }
\label{fig4}
\end{figure*}

  Before investigating  composite diagrams we have checked the correlation between magnitudes  of similar wave-bands in different catalogs. 
   Doing this our main purpose was to verify the  efficiency of  composite color-color diagrams. 
  For this reason and  taking into account the lower variability amplitude in the far IR,  we have  considered all the  variables of  the BIS catalog, including those not present in G17.  
 The results  shown in Fig.~\ref{fig4} indicate a general good agreement  between $AKARI$, $IRAS$, and $WISE$,  irrespective of the spectral types. 
The good correlation along the whole range of magnitudes  between $AKARI$ and $IRAS$ is also  due to the lack of data points in the $AKARI$  ~catalog for faint stars.  Many of these last  stars are present in $IRAS$ though with low quality factors, 
resulting in a worse correlation with $WISE$  which has a more extended and accurate data set.
 
%
%
The composite color-color diagrams derived from similar passbands show statistically a good correlation, with few discrepant cases. 

The IRAS [12]$-$[25] and $WISE$ 12$-$22 ($W3-W4$) colors show an overall satisfactory correlation in spite of the spread of 0.6~mag in the IRAS color of the non-variable stars while the same stars are grouped in just 0.2 mag in the $WISE$ color (see x axis of Fig.~\ref{fig3} and Fig.~\ref{fig2}).
The two most discrepant cases are the large amplitude, irregular variables BIS\,207 and 198, 0.3~mag out of the ridge line.

 The number of BIS sources  present in the $AKARI$~  database is lower than in other databases and for these stars only the fluxes at 9 and 18~$\mu$m  are available. In spite if this, the $AKARI$~   fluxes can be used in composite diagrams that equally give  interesting results.
The $AKARI$~  color S09$-$L18 correlates very well with the similar IRAS [12]-[25] color (r$=$0.80); only BIS\,255 is definitely out (0.6~mag) of the correlation. This star however is not peculiar in the IRAS [25]-[60] {\it vs} [12]-[25] plot (Fig.~\ref{fig3}) and is not variable.

The correlation of $AKARI$~  S09$-$L18 with the $WISE$ 12$-$22 color is even better (r$=$0.90), again with BIS\,255 standing out (0.6~mag): in this plot also BIS\,007 (which  anyhow  is a Mira$-$type star) is markedly out (0.3~mag) of the correlation; both stars have good quality flag in the $AKARI$~  catalog and are positioned near to the ridge line in the 12$-$22 ($W3-W4$) {\it vs} 4.6$-$12 ($W2-W3$) $WISE$ color$-$color plot.  

\begin{figure}[t]
\centering
\includegraphics[width=12cm,angle=0]{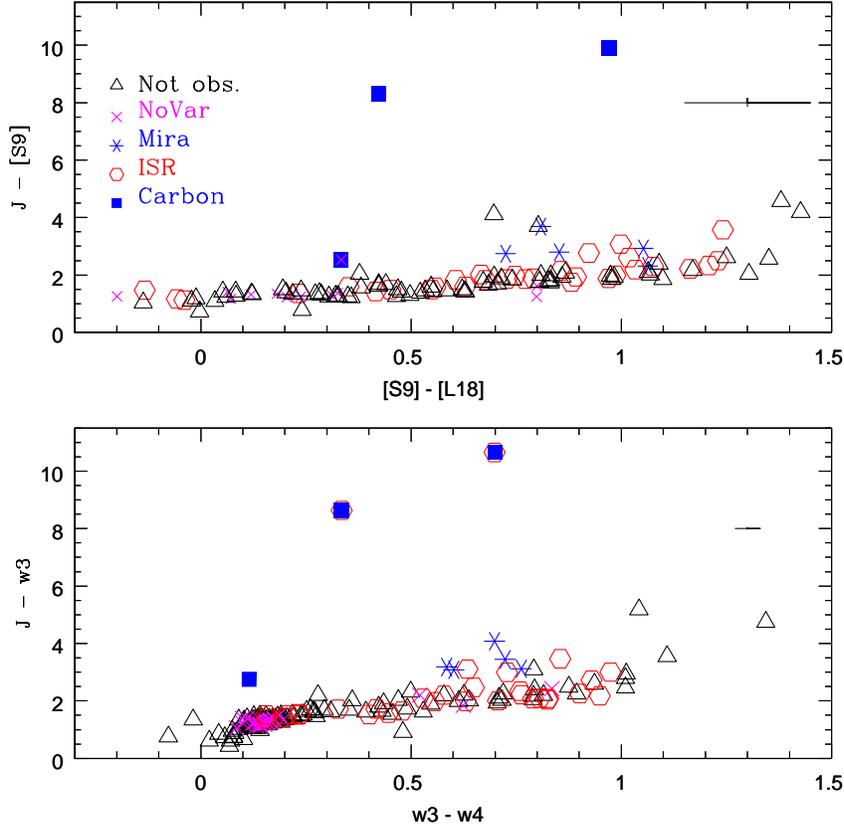}
\caption{ Top: $J$-[S09]   vs [S09]-[L18] ~color$-$color diagram. 
The errors in the x axis are typically 0.15 mag.; in the y axis the errors are due to the [S09] magnitude and are similar, but  not relevant thanks to the extended magnitude range.
 ~~Bottom: $J$-$W3$   vs $W3$-$W4$   diagram.  The errors  are  typically 0.05 mag. These values are indicated by the crosses at the corners. 
  }
\label{fig5}
\end{figure}

 Important diagrams  applied to different  passbands have been constructed and accurately studied using the $AKARI$ ~database by \citet{ishi11} and  \citet{ita10}; we have successfully  applied the same combinations to our data. In  Figure 3 of \citet{rossi16} we have plotted  the  combination proposed by  \citet{ishi11}  $K$-[S09] vs [S09]-[L18], that we have compared with K-$W3$ vs $W3$-$W4$.   
Using J instead of K  yields to  similar results with even more extended range in the y axis
as shown in   Fig.~\ref{fig5} of the present work.
Again obscured carbons and oxygen rich stars are distributed along well separated paths and  oxygen Miras lie in the upper part of the strip of variables.
In the top panel the range of the sources is spread along the x axis more than in the bottom panel; for this reason [S09]-[L18] is not too effective in separating variable from non variable stars.
    %
%

 \smallskip


 \section{Concluding remarks}   \label{Conc}

 
In the present paper we  have described  the distribution of the BIS catalog stars in several color$-$color IR plots. We used the spectral classification and variability classes derived in our companion paper G17 for a subsample of BIS stars to interpret the distributions of these infrared plots.

\textbullet~We used the classical $J-H$ {\it vs} $H-K$ diagram to check the luminosity class and to look for the presence of cool dusty envelopes around the stars:

\textbullet~All the diagrams involving $WISE$ colors are efficient in separating the early-M type, non$-$variables, from the mid- and late- type stars (all variables). 
The presence of emission lines in the optical spectrum is not discriminant for the position of the stars in any of  the $WISE$ diagrams. 

It is worth noticing that all the confirmed Miras have the $W1-W2$ color greater/equal than 0.5 mag, while the other (non dusty) stars lie below this limit.
The dust enshrouded carbon stars are always outstanding in all plots, aligned in a diagonal path: naked carbon star BIS\,194 falls in the group of the early O$-$rich stars and cannot be discriminated with these diagrams. 
The stable stars are clustered in the lower left (blue) corner of all plots.

 \textbullet~ 
  Similar results have also been obtained from  composite color-color diagrams. the agreement holds for optically variables too;
  this result is  supported  by the   satisfactory correlation  found  comparing   data  from
   different experiments.  Carbon stars are always well distinguished from M  stars; more remarkable is that  the -few- confirmed Miras  in every diagram  are slightly but  clearly separated above the strip defined by the other variables.
    
\textbullet~We have  analysed our sample in the  IRAS [25]$-$[60] {\it vs} [12]$-$[25]  color$-$color plot (Fig.~\ref{fig3}) finding only an overall general agreement with the  zones defined by \citet{veen88}. 
While the carbon and  stable stars  lie in the expected regions,   the irregular, large amplitude  variables of M types are distributed  everywhere in the diagram. Strong discrepancies in the y axis might be  attributed to the large uncertainties of the fluxes at 60 $\mu$, but for the same stars  less obvious are the locations in the x axis where the quality factors are generally good (3, 2). 

\medskip

   From our new  optical spectra and  from the infrared data  we have been able to better define the   population of the BIS catalog.
 Among  the  "normal" stars of the halo population, we have  also discovered   few objects deserving special attention.
 
 An outlier  is BIS\,010.  The optical magnitudes of this  optically normal M5 star  are probably  wrong in every catalog derived from the POSS, being the northern component of an apparent close binary  : we took a spectrum of the companion  finding  a  G   star, 
  extremely faint in the R band   disappearing already  in the  near infrared images (see G17).
The peculiarity of BIS\,010 is the extreme infrared brightness revealed by the $WISE$, IRAS and $AKARI$~  coherent magnitudes that put this star very far from the bulk in every color-color diagram, out of the scale of the plots, presented in this paper, but well separated from the strip of the dusty carbon stars. 
In the near infrared  color-color diagram the red component of  BIS\,010 is not peculiar  with  $J-H$=0.94, $H-K$=0.40, while  the $WISE$ magnitudes are  $W1$ = 10.3, $W2$ = 8.58, $W3$ = 4.34,  $W4$ = 1.63). 
 
 Another extremely puzzling object is  BIS\,131, a K2 star whose peculiarity  in  the optical spectrum   is the presence of  permitted and  forbidden  emission lines  \citep{Ros10}.
 Our long term photometric monitoring is revealing  slow, small amplitude photometric variability and  minor spectroscopic variations of the emission lines (not yet  published).
Similarly to BIS\,010  the colors of  BIS\,131 are normal in the optical and near  IR, and extreme
 in the  far IR.
 
 \medskip
 Combining the optical  information from G17  with the infrared ones described in the present paper
  we are now in the position to  make statistical predictions on the overall stellar content of the whole BIS sample. 
Among the  181 BIS sources still not studied in the optical,  we expect that 15\%  are earlier than M2 while the rest are late M  with few dusty M-type; a few Miras (one is already known, BIS\,176) may be present. No dusty Carbon star is expected, while a few CH or R Carbon stars may be still to be present.

\bigskip
\bigskip

\begin{acknowledgements}
{\bf Acknowledgements.}
This research has made use of the SIMBAD database, operated at CDS, Strasbourg, France.
This publication has made use of data products from:
 the Two Micron All$-$Sky Survey database, which is a joint project of the University Massachusetts and the Infrared Processing and Analysis Center/California Institute of Technology;
the Wide$-$field Infrared Survey Explorer, which is a joint project of the University of California, Los Angeles, and the Jet Propulsion Laboratory/California Institute of Technology, funded by the National Aeronautics and Space Administration; 
the Northern Sky Variability Survey (NSVS) created jointly by the Los Alamos National Laboratory and University of Michigan;
 the NASA/IPAC Extragalactic Database (NED) which is operated by the Jet Propulsion Laboratory (JPL), California Institute of Technology, under contract with the National Aeronautics and Space 
Administration.
\end{acknowledgements}

\end{document}